\documentclass[twocolumn,nobibnotes,showpacs,showpacs,preprintnumbers,amsmath,amssymb,superscriptaddress,]{revtex4}
\usepackage{graphicx}% Include figure files
\usepackage{dcolumn}% Align table columns on decimal point
\usepackage{bm}% bold math
\usepackage{lineno}
\begin{document}

\preprint{}

\title{Resonance and frequency-locking phenomena in spatially
extended phytoplankton-zooplankton system with additive noise
and periodic forces}
\author{Quan-Xing Liu}
\email{liuqx315@sina.com}
\affiliation{Department of Mathematics, North University of China,\\
Taiyuan, Shan'xi 030051, People's Republic of China
}%

\author{Bai-Lian Li}%
\email{bai-lian.li@ucr.edu}
\affiliation{Ecological Complexity and
Modeling Laboratory, Department of Botany and Plant Sciences,
University of California, Riverside, CA 92521-0124, USA}%

\author{Jin Zhen}
\altaffiliation[Corresponding author]{}
\email{jinzhn@263.net}
\affiliation{Department of Mathematics, North University of China,\\
Taiyuan, Shan'xi 030051, People's Republic of China
}%
\date{\today}% It is always \today, today,

\begin{abstract}
It is shown that natural systems are undeniable subject to random
fluctuations, arising from either environmental variability or
internal effects. In this paper, we present a spatial version of
phytoplankton-zooplankton model that includes some important factors
such as external periodic forces, noise, and diffusion processes.
The spatially extended phytoplankton-zooplankton system is from the original study by
Scheffer [M Scheffer, Fish and nutrients interplay determines algal
biomass: a minimal model, Oikos \textbf{62} (1991) 271-282].  Our results show that
the spatially extended system exhibit
a resonant patterns and frequency-locking phenomena. The system also shows that
the noise and the external periodic forces play a constructive role
in the Scheffer's model: first, the noise can enhance the
oscillation of phytoplankton species' density and format a large
clusters in the space when the noise intensity is within certain interval.
Second, the external periodic forces can induce 4:1
and 1:1 frequency-locking and spatially homogeneous oscillation
phenomena to appear. Finally, the resonant patterns are observed
in the system when the spatial noises and external periodic forces
are both turned on. Moreover, we found that the 4:1 frequency-locking
transform into 1:1 frequency-locking when the noise intensity increased. In addition to
elucidating our results outside the domain of Turing
instability, we provide further analysis of Turing linear stability
with the help of the numerical calculation by using the Maple
software. Significantly, oscillations are enhanced in the system
when the noise term presents. These results indicate that the
oceanic plankton bloom may partly due to interplay between the
stochastic factors and external forces instead of deterministic
factors. These results also may help us to understand the effects arising
from undeniable subject to random fluctuations in oceanic plankton
bloom.
\end{abstract}


%Use showkeys class option if keyword
\maketitle

\tableofcontents
\section{Introduction}

\begin{figure}
\includegraphics[angle=0,scale=0.3]{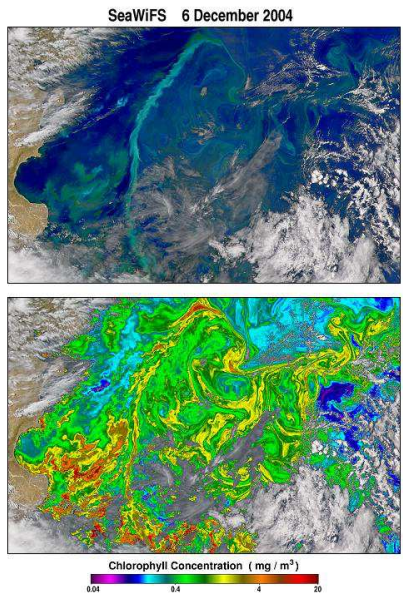}
%(b)\includegraphics[angle=90,scale=0.2]{picture/satelitte2.ps}
\caption{ (Color online) This satellite images are the field
observation of the phytoplankton blooms in the Malvinas current
region. The enhanced natural color images show actual differences in
water color while the pseudocolor image shows chlorophyll
concentration. The images are taken form
http://oceancolor.gsfc.nasa.gov, with permission from Janet W.
Campbell.}\label{fig1}
\end{figure}

Many mechanisms of the spatiotemporal variability of natural
plankton populations are not known yet. Pronounced physical
patterns, such as thermoclines, upwelling, fronts, and eddies, often
set the frame for the biological process. While many parameters and
variables can be learned from standard oceanographic measurements of
temperature, salinity, nutrients, and biomass concentrations of
phytoplankton and zooplankton, now new technologies such as remote
sensing are being developed for observing the ecosystem.
Measurements of the underwater light field are made with
state-of-the-art instruments and used to calculate concentrations of
phytoplankton biomass, such as chlorophyll, as well as other forms
of organic matter. More recently, satellite remote sensing and
detailed numerical simulations identify some spatial patterns
including filaments, irregular patches, sharp gradients, and other
complex structures involving a wide range of spatial scales and time
scales by utilizing the species'
concentration~\cite{report1,bettencourt-2007-in-situ,abraham:373}.
Figure~\ref{fig1} shows two pictures deriving from the field
observation by the satellites remote sensing, where color gives us
very useful information and ideas about the changes of chlorophyll
concentrations in the two-dimensional spaces. The more pictures
about the spatial patterns of the phytoplankton can be obtained from
the web
\href{http://oceancolor.gsfc.nasa.gov}{http://oceancolor.gsfc.nasa.gov},
such as the stripelike, spotted-like, clockwise-rotating  (or
counterclockwise-rotating) spiral waves, spatial chaotic patterns,
etc. Especially, In Ref.~\cite{report1} from an observing system for
monitoring the ecosystem is approached by posing five questions: How
is the ecosystem changing? What are the forcing factors causing it
to change? How does the ecosystem respond to natural and human
forces? Can we predict future changes? And what are the consequences
for stakeholders in our region?

Learning from historical records and from our field observations in
Refs.~\cite{Durbin,May2003}, one knows that the timing and magnitude
of phytoplankton blooms vary significantly on interannual to longer
timescales. For example, in some years, e.g., 2002, the spring bloom
phytoplankton occurred in April, coinciding with vernal flowering on
land, whereas in other years, e.g., 1999, the bloom occurred in
February. Although there are only a handful of dynamics states that
phytoplankton populations may exhibit stable equilibrium,
deterministic extinction, stable population oscillations, and
irregular fluctuations, there is a long list of factors that may
interact to determine the community dynamics, e.g, competition,
predation, parasitism, mutualism, age, stage and genetic structure,
spatial structure of the habitat, climate, physical and chemical
parameters. Thus, a key step in analyzing the community's dynamics
is to untangle these mixtures of interacting factors and to identify
their essential for the observed dynamics. Especially, from the
experimental plankton communities~\cite{Desharnais} [rotifer-algal]
one knows that: first, the age structure of the predator populations
is necessary to generate qualitatively correct predictions of
population dynamics such as stability versus cycles. Second, the
rapid evolution of population, including the alga and rotifers, is
regarded as critical processes occurring on the same time scale in
the ecological dynamics, and learned that microcosms may not just
serve as a means to check model's assumptions, but that the results
of microcosm studies can lead to novel insights into the function of
biological communities. From
Refs.~\cite{E.E.Popova10011997,EckartSteffen,HorstMalchow}, we know
that the interannual variation in zooplankton and phytoplankton
species might be the result of changes in climate, and the
photosynthetic of the phytoplankton growth strongly depends on the
intensity of the light. Hence, these period factors are regarded as
the external periodic forces in the plankton systems. The variances
of species evolving in time and space may be well understood.

Besides these periodic factors, there are many other stochastically
factors causing phytoplankton-zooplankton blooms to transit
extinction. For example, the effects arise from rivers' pollution on the
phytoplankton-zooplankton ecosystem is one of the ways that humans
affect the marine ecosystem. Large rivers are major mechanisms for
nutrient delivery to the ocean. Hence, the quality of river water
affects freshwater ecosystems and oceanic food webs. Away from this,
the long term climatic variation also shows the stochastic factor
for the ecological system [see~\cite{benzi-2007} for giving a review
about this points]. In the phytoplankton-zooplankton systems, the
random fluctuations also are undeniably arising from either
environmental variability or internal species. To quantify the
relationship between fluctuations and species' concentration with
spatial degrees of freedom, the consideration of these fluctuations
supposes to deal with noisy quantities whose variance might at times
be a sizable fraction of their mean levels. For example, the birth and death
processes of individuals are intrinsically
fluctuations~\cite{Malchow2004}, which become especially pronounced
when the amount of individuals is small. The interaction between the
oceanic zooplankton and fish, which are far from being uniformly
distributed, also introduce randomness~\cite{Vilar}. In addition,
these processes can be regarded as parameter fluctuates irregularly with spaces and time.
For example, phytoplankton production is effected by iron
concentration, which can be elevated in surface water after
rain~\cite{KennethS.Johnson04192002}. These unavoidable fluctuations
can also interact with the system's nonlinearities to render
counterintuitive behavior, in which an increase in the noise level
produces a more regular behavior~\cite{sagues:829}. Now, it is
natural to ask what are spatiotemporal behaviors of the spatially extended plankton
system if these external periodic forces and the irregular
fluctuation factors work on. Especially, can the periodic
oscillation retain to appear? In order to understand the
stochastically force, we give the short introduction of the noise
application in the physics and biology.

Recently, research interest has shifted to the effects of noise in
spatially extended system~\cite{Garca-Ojalvo, sagues:829}.
Well-known examples in zero-dimensional (ODE) systems are noise-induced
transition~\cite{Horsthemke}, stochastic
resonance~\cite{Gammaitoni,PhysRevLett.74.2130,RevModPhys.70.223},
enhanced spatial synchronization~\cite{PhysRevLett.83.4896}, and
noise-sustained
oscillations~\cite{PhysRevE.61.R1001,PhysRevLett.87.098101,ChangsongZhou}
[the reader can read an overview written by F Sagu\'{e}s in
Ref.~\cite{sagues:829}]. More recently, effects of noise in
spatially distributed systems, including noise-induced pattern
formation~\cite{PhysRevE.67.021113,wang:036210,sagues:829},
noise-induced fronts~\cite{PhysRevE.59.98,PhysRevLett.88.138301},
etc, have been extended study by many communities. In these and
other noise-related phenomena, multiplicative noise, which couples
to the system state plays a very special role. However, prominent
effect has been also found for additive noise. Such influence has
been observed in noise-induced pattern
formation~\cite{wang:036210,PhysRevE.58.4355,Riaz,PhysRevLett.76.2609,PhysRevLett.74.2130}.
A recent report~\cite{sailer:056209} demonstrated that additive
noise which globally alternates between with two different
monostable excitable dynamics yields pattern formation. Meanwhile,
in recent years several theoretical investigations have been done on
noise-induced effects in population dynamics,
see~\cite{Spagnolo,CarlZimmer04021999,OttarN.Bjornstad07272001,PhysRevE.54.706,PhysRevLett.80.4099,
Giardina,Spagnolo2002,Valenti2004}. Noise, such as that from the
natural variability, are inevitably presented in these type of systems,
but its effects have not yet been addressed completely. In this
paper, we report resonant patterns and frequency locked oscillations
induced by additive noise and external forces in
phytoplankton-zooplankton systems, and take into account the
interplay among noise, external forces, and diffusion processes. In
the following, we first give the spatially extended model and method we used, and then
the describe our results.

\section{Model}

From the recent perspective of Pascual~\cite{MercedesPascual2005}, the
physical environments play an important role in the biota. If the
climate sharply changes, the population abundance will change.
Especially, the spatiotemporal dynamics of global population
abundance, such as aggregated over the whole space, can be
approximated by mean-field-type equations in which the functional
forms specifying the growth rates and interactions have been
modified as power functions. And, the effect of interactions on
local (or individual scales) can be represented implicitly by the
changed form of the functions to describe interactions on global
scales. We want to focus on the effects of a periodic varying and
the stochastic fluctuation factors to the spatially extended phytoplankton and
zooplankton model. Following Scheffer's minimal
approach~\cite{Scheffer1991a} and the previous
analysis~\cite{PhysRevE.64.021915,medvinsky:311,Malchow1993,MercedesPascual},
we study a two-variable phytoplankton and zooplankton spatial model
involving time-periodic forces and the fluctuation term to
describe the influence of spatial noisea spatial on pattern formation.
 The spatially extended model is written as
\begin{widetext}
\begin{subequations}
\label{eq:II2}
\begin{equation}
\frac{\partial p}{\partial t}=r
   p(1-p)-\frac{ap}{1+bp}h+A\sin(\omega t)+d_{p}\nabla^2 p
 ,\label{subeq:II2a}
\end{equation}
\begin{eqnarray}
\frac{\partial h}{\partial
t}=\frac{ap}{1+bp}h-mh-f\frac{nh^2}{n^2+h^2}+A\sin(\omega
t)+\eta(\bm r,t)+d_{h}\nabla^2
   h,\label{subeq:II2b}
\end{eqnarray}
\end{subequations}
\end{widetext}
where the parameters $r$, $a$, $b$, $m$, $n$, $d_{p}$, $d_{h}$, and
$f$ refer to the previous
works~\cite{PhysRevE.64.021915,medvinsky:311} on the dimensionless
model~\eqref{eq:II2} absence of periodic force and noise term. Here
$p(x,y,t)$ and $h(x,y,t)$ are scalar fields representing the
concentrations of phytoplankton species and zooplankton species in
the two-dimensional space. It must also be kept mind that the individuals of the
population have not been at the same point in space at previous
times, so they are a function of time. The periodic force is
assumed to be sinusoidal with amplitude $A$ and frequency $\omega$.
The periodic force is considered as additive version for reason that
toxins produced by different phytoplankton species has a significant
role in shaping the dynamical behavior of marine plankton
ecosystems~\cite{Kirk1992}. The zooplankton population tries to
avoid the areas where the concentration of phytoplankton is very
large. The reason may be either dense concentration or the effect of
toxic substance released by phytoplankton, of course including the
others factor as introduced in previous section, such as human
action, nature sharp changing-red tides and localized outbreaks and
occur in coastal water and fronts~\cite{Truscott1992}. Buskey and
Stockwell~\cite{Buskey1993} have shown in their field study that
macro- and meso-zooplankton populations are reduced during the
blooms of Chrysophyte (Aureococcus anophagefferens). These factors
are not directly correlated with species internal parameters, so
that the $A \sin(\omega t)$ be added in the equation instead of
$Ap(1-p)\sin(\omega t)$. Such consideration is meaningful in the
biology systems (see Ref.~\cite{Petrovskii2001,medvinsky:311}). In
Eq.~\eqref{eq:II2}, the stochastic factors are taken into account as
the term, $\eta(\bm r,t)$. Of course, due to the coupling, the noise in
Eq.~\eqref{subeq:II2b} will have its influence on
Eq.~\eqref{subeq:II2a} as well.
The noise term
$\eta(\bm r,t)$ is introduced additively in space and time, which is
the Ornstein-Uhlenbech process that obeys the following stochastic
partial differential equation~\cite{report2}:
\begin{equation}
\frac{\partial \eta(\bm r,t)}{\partial t}=-\frac{1}{\tau}\eta(\bm
r,t)+\frac{1}{\tau}\xi(\bm r,t), \label{eq:noise}
\end{equation}
where $\xi(\bm r,t)$ is a Gaussian white noise with zero mean and
correlation,
\begin{equation}
\langle\xi(\bm r,t)\xi(\bm r',t') \rangle =2\varepsilon \delta(\bm
r-\bm r')\delta(t-t'). \label{eq:noise_1}
\end{equation}
The colored noise $\eta(\bm r,t)$, which is temporally correlated
and white in space, satisfies
\begin{equation}
\langle\eta(\bm r,t)\eta(\bm r',t') \rangle
=\frac{\varepsilon}{\tau} \exp\Big(-\frac{|t-t'|}{\tau}\Big)
\delta(\bm r-\bm r'), \label{eq:noise_2}
\end{equation}
where $\tau$ controls the temporal correlation, and $\varepsilon$
measures the noise intensity.

In this work, we rely on numerical simulations of the model of
Eqs.~\eqref{eq:II2} and~\eqref{eq:noise}.
We here consider spatiotemporal evolution of this system with space
white noise and colored noise evolving in time when the system lies
within the regime of self-sustained Hopf oscillation. Note that the explicit
form of the noise term represents only on the fluctuating recruitment
rate (or death rate) has been studied by Malchow et
al~\cite{Malchow2004}, in which the parameter $m$ only is regarded as
noise form, i.e., $m=m_0+\xi(t)$, $\xi(t)$ denotes Gaussian white
noise. For the absence of the external periodic
forces and colored noise, Hopf instability occurs and spatially homogeneous oscillation comes up
when the parameter $f$ less than
the critical value $f_{H}=0.3398$ [see Appendix A] whose value depends
on the other parameters.  Moreover, from previous analysis and
numerical simulations~\cite{garvie:775,liuquanxing} show that spiral
waves structure exists in the two-dimensional space and the
spatiotemporal chaos will emerge through its far-field breakup when
the parameter $f$ within the mixed domain of the Hopf-Turing
instability, but the parameters are outside the Turing instability
domain in present paper [see Appendix A for a simple analysis].
 In fact, from recent perspective that the 2:1 resonance phenomena
 on the spiral wave may will appear when the system undergos saddle-node
  or Hopf bifurcation~\cite{sandstede:494}. In other words, the temporal
  period of the bifurcation patterns is twice the period of the primary
  spiral, i.e., the Hopf frequency $\omega_{H}$ needs to be in a 2:1
  resonance with the rotation frequency of the spiral wave. However,
  here we show that the system also exhibits spatiotemporal chaos
patterns and the 4:1 and 1:1 resonances with the period of the external
 forces
when fluctuations and periodic forces are
considered. Except when it explicitly is pointed out, we take
parameters $r=5$, $a=5$, $b=5$, $m=0.6$, $n=0.4$, $f=0.3$,
$d_p=0.05$, and $d_h=0.5$ throughout this paper. These parameters
estimate by Medvinsky et al~\cite{PhysRevE.64.021915,medvinsky:311,Malchow1993}. From these
References, we know that these
parameters are meaningful from the ecological point of view. The noise intensity
$\varepsilon$ and correlation time $\tau$ are adjusted as control
parameters.

\section{Results}

Extensive testing was performed through numerical simulations to the
described model~\eqref{eq:II2} and the results are shown
this section. In simulation, zero-flux boundary conditions are used
and time step $\Delta t=0.05$ time unit. The space step $\Delta
x=\Delta y=1$ length unit and the grid sizes in the evolutional
simulations are $N\times N$ ($N=200$). The Fourier transform method
is used for the deterministic part in Eq.~\eqref{eq:II2}. On the
discrete square lattices, the stochastic partial differential
Eq.~\eqref{eq:noise} is integrated numerically by applying the Euler
method. Several different discrete methods (simple Euler,
Runge-Kutta, and Fourier transform) was checked, and the results
indicate that Fourier transform accurately approximates solutions of
Eq.~\eqref{eq:II2}. On the other hand, the Fourier method offers the
speed advantage over other numerical methods. We find that on the PC
computer the Fourier method runs about 3-4 times faster than the
Euler integration using the same time step and space step. Code is
implemented in Matlab 7.3 and the \emph{fft2}, \emph{fftshift},
\emph{ifft2}, and \emph{ifftshift} functions were used for the main
numerical integration.

Although the noisy fluctuations may sometimes causing the variable
($p$ and $h$) to be less than zero, it will lead to the
diffusion-reaction system with cutoff effect at low densities when
the species extinction is taken into account
explicitly~\cite{Petrovskii2004}. According to the spatially
extended model~\eqref{eq:II2}, at each position in space, whenever
the population densities fall below certain prescribed value
$\varepsilon$ they are set to zero or sufficient small positive
constant~\cite{Petrovskii2004,Wilson}. From the biological point of
view, in this paper we set that they equal to 0.0001 when the
variables change to negative. Note that we are not much concerned
here with the ``exact`` value of $\varepsilon$, for the reason that
an attempt to estimate the ``exact'' value would hardly make any
ecological sense in terms of very schematic model~\eqref{eq:II2}. To
compare with the numerical results under the different cases, we
used the same initial conditions that are randomly perturbed (the
perturbations are space-independent) by homogenous equilibrium
$(p^{*},h^{*})=(0.3944,1.7998)$, except when they are explicitly
pointed out.

\subsection{Dynamics under the presence of noise only}

\begin{figure}[h]
\includegraphics[angle=0,width=10cm]{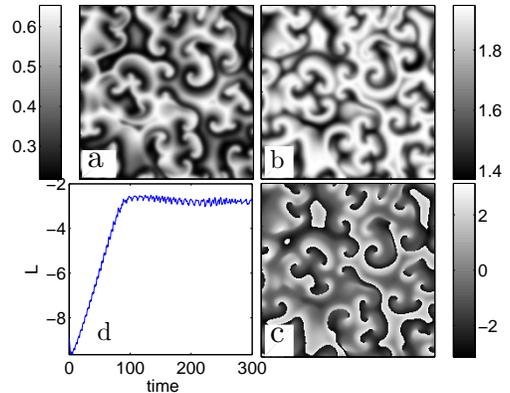}
\caption{A typical grey-scaled snapshots of spatiotemporal chaotic
patterns at $t=300$, when the system~\eqref{eq:II2} evolves in time
without noise term. (a) and (b) the spatial patterns of $p$ and $h$,
respectively; (c) the phase angle, $\theta$ (see the definition in
text); (d) the amplitude, $\mathcal{L}$ evolves in time. [see the
\emph{Movie-1}, and additional movies format available from
Ref.~\cite{movies}]} \label{nonoise}
\end{figure}

\begin{figure}[h]
\includegraphics[angle=0,width=10cm]{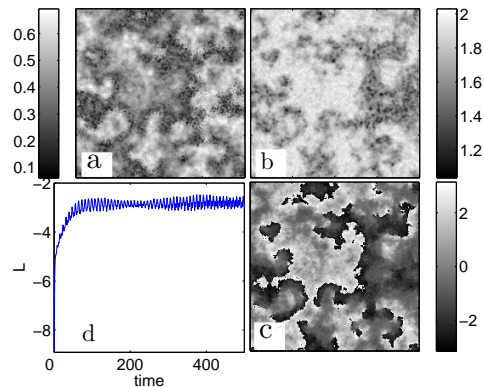}
\caption{ A typical grey-scaled snapshots of spatiotemporal chaotic
patterns at $t=300$, when the system~\eqref{eq:II2} evolves in time
with noise term. The meanings of the images are correspondent to the
Fig.~\ref{nonoise}, with the values of the parameters are $\tau=1$
and $\varepsilon=0.001$. [see the \emph{Movie-2}, and additional
movies format available from Ref.~\cite{movies}]} \label{onlynoise1}
\end{figure}

\begin{figure}[h]
\includegraphics[angle=0,width=10cm]{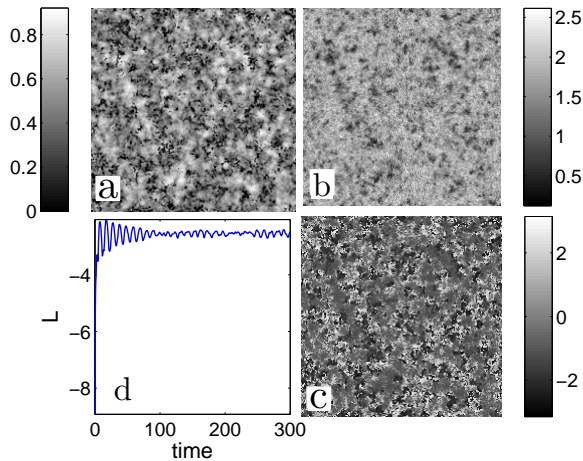}
\caption{A typical grey-scaled snapshots of spatiotemporal chaotic
patterns at $t=300$. Same situation as the Fig.~\ref{onlynoise1} but
$\varepsilon=0.05$. [see the \emph{Movie-3}, and additional movies
format available from Ref.~\cite{movies}]} \label{onlynoise2}
\end{figure}

\begin{figure}[h]
\includegraphics[angle=0,width=7cm]{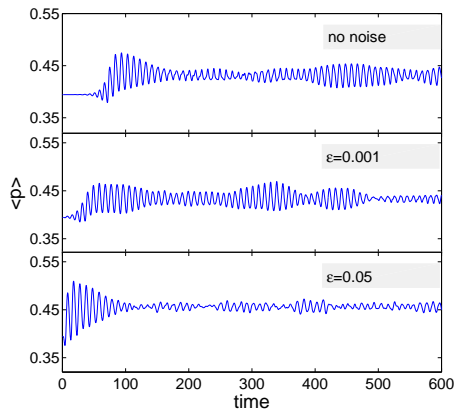}
\caption{Directly compare with the mean values of variable $p$
without and with noise term. Same situation as the
Fig.~\ref{nonoise}-\ref{onlynoise2}.} \label{addfig1}
\end{figure}

In the ecological systems, noise-sustained and noise-induced spatial
pattern formations have been discussed  in recent
years~\cite{CarlZimmer04021999,OttarN.Bjornstad07272001,PhysRevE.54.706,PhysRevLett.80.4099,Giardina,Spagnolo2002,Valenti2004}.
For the convenience of discussion about the resonant pattern
formation induced by additive noise and the external forces, here we
first present a brief description about the affection arising from
the fluctuations in the system~\eqref{eq:II2}. From our results show
that the sole noise also plays a constructive role in the
model~\eqref{eq:II2} by maintaining (or eliminate) the large spatial
clusters [cf. Fig.~\ref{onlynoise1}, \ref{onlynoise2} and partly
\emph{movie-1},\emph{movie-2}, \emph{movie-3}] and enhancing the
aperiodic oscillation [see Fig.~\ref{addfig1}]. Some typical
snapshots of the spatial patterns at $t=300$ presented in
Fig.~\ref{nonoise}, and Fig.~\ref{onlynoise1}, \ref{onlynoise2},
where before and after the noise is added, respectively [see caption
of Figures]. The ratio's distribution of two species is a key
measurement in the ecological field. To characterize this
measurement in the system~\eqref{eq:II2}, we introduce the
parameter, $\theta$ (the $\theta \in [-2\pi, 2\pi]$) and define as
following:
\begin{equation}
\theta=\arg((p_{s,k}-p^{*})+i(h_{s,k}-h^{*})), \label{eq:arg}
\end{equation}
where $s, k=1, 2, \cdots, N$. $\theta$ indicates the relation between the two variables of the system,
called the phase angle. Note that it also depicts the variance of
the amplitude relevant the two species in the space, for the reason
that its variables are the difference around the stable points,
$p^{*}$ and $h^{*}$. Similarly, the following parameter,
$\mathcal{L}$, depicts the average amplitude for the system in the
space. However, we find that it is equivalent to the mean value of
the variables in this system [cf.
Fig.~\ref{nonoise},\ref{noiseandforce1},\ref{noiseandforce2}, and
Fig.~\ref{addfig1}], so we only use it in this section.
\begin{equation}
\mathcal{L}=\ln (\frac{\sqrt{\sum(p_{s,k}-p^{*})^2}}{N}),
\label{eq:amp}
\end{equation}
where $s, k=1, 2, \cdots, N$. The spatiotemporal chaotic spiral
patterns appears when the noises are free [see Fig.\ref{nonoise}],
and the large clusters can be recognized in the space. The
$\mathcal{L}$ or $<p>$ (the $<p>$ denotes the spatial average of $p$
rather than the time average) oscillates with small fluctuations
(quasi-period)~\cite{misc10}, as shown in Fig.\ref{nonoise}(d) (and
Fig.~\ref{addfig1}), in which the abscissa is time and the ordinate
is variable, $\mathcal{L}$. The behavior of the system undergoes
drastic changes when the noises are turned on (cf.
Fig.~\ref{nonoise}, \ref{onlynoise1}, and \ref{onlynoise2}). First,
the large spatiotemporal chaotic clusters die out gradually with the
increasing of noise intensity; second, the oscillations of
$\mathcal{L}$ become more obvious when noise intensity is within a
certain regions, but it stays at a fixed value again if the noise
intensity is further increasing. In sum, compared with
Fig.~\ref{nonoise}, \ref{onlynoise1}, and \ref{onlynoise2}, it
should be noted that the noise plays a constructive role in the
relation between phytoplankton and zooplankton in the space.


\subsection{Dynamics under the presence of the external forces only}

From the previous analysis~\cite{EckartSteffen}, it is known that
the plankton system~\eqref{eq:II2} has a very complex dynamical
behaviors such as the scenario of bifurcations and chaos  when the
system was not considered the spatial degree of free. However, we
here consider the case that the parameter values are within the Hopf
oscillation interval and the spatial degree of free was taken into
account.

The parameter values under we take in previous Section II, and also
the system is noise free in this subsection, thus the spatially
homogeneous oscillation does not respond to the external periodic
force if the amplitude $A$ is below a threshold $A_{c}$, whose value
depends on the external period $T_{in}=\frac{2\pi}{\omega}$. Above
the threshold, the system~\eqref{eq:II2} produces oscillations about
period $T_{out}$ with respect to external period $T_{in}$, this
phenomena are called frequency locking or resonant response, i.e.,
when the system produces one spike within each of the $M$ ($M=1, 2,
3, \cdots$) periods of the external force, that is, $M:1$ resonant
response. In present paper, the output period $T_{out}$ is defined
as follows: $T_{i}$ is the time interval between the $i$th spike and
the $i+1$th spike. $q$ spikes are taken into account and the average
value of them is $T_{ out}$, where $T_{
out}=\frac{\sum\limits_{i=1}^{q} T_{i}}{q-1}$~\cite{sifenni}.

\begin{figure}[h]
\includegraphics[angle=0,width=6cm]{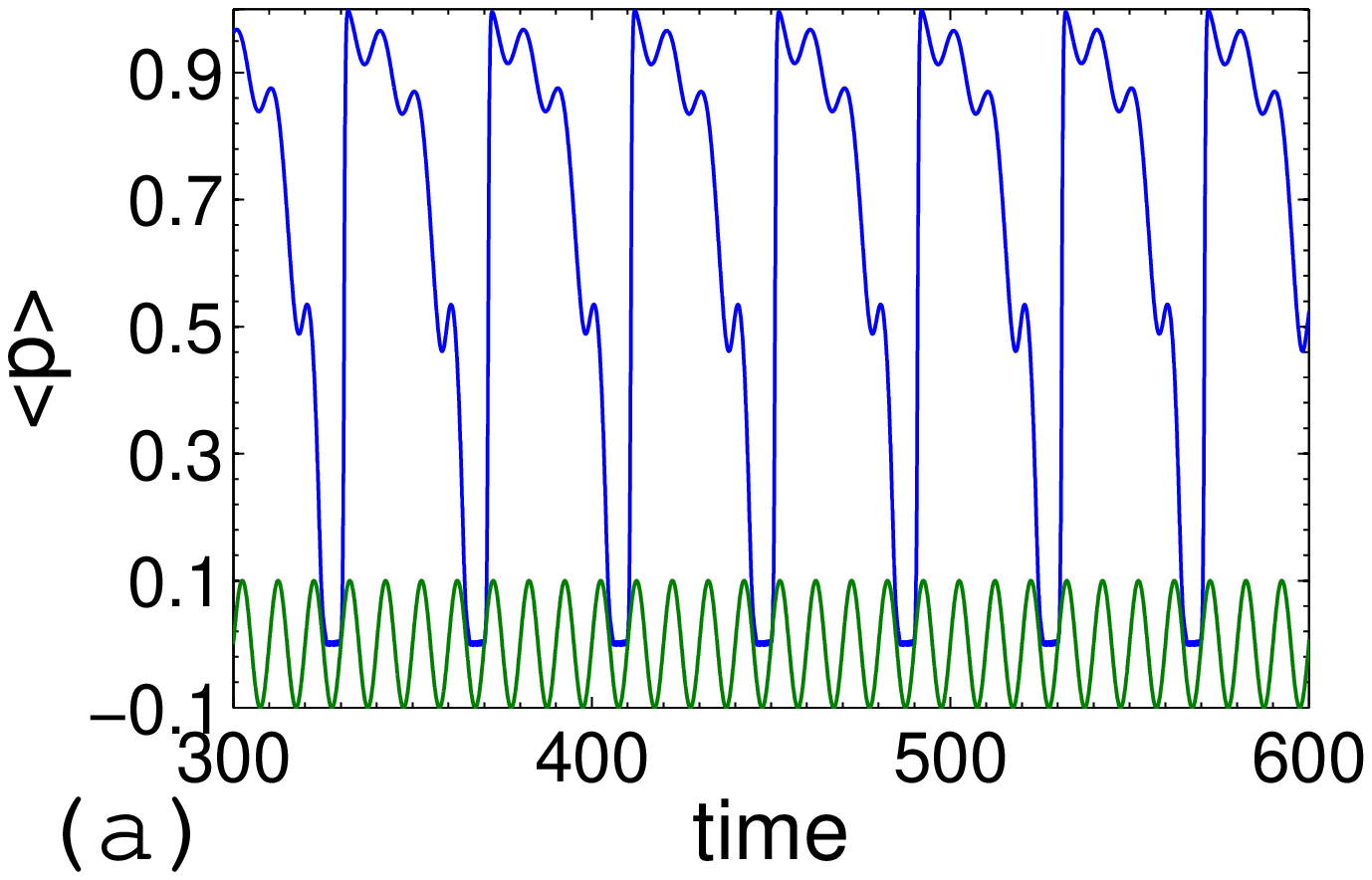}\\
\includegraphics[angle=0,width=6cm]{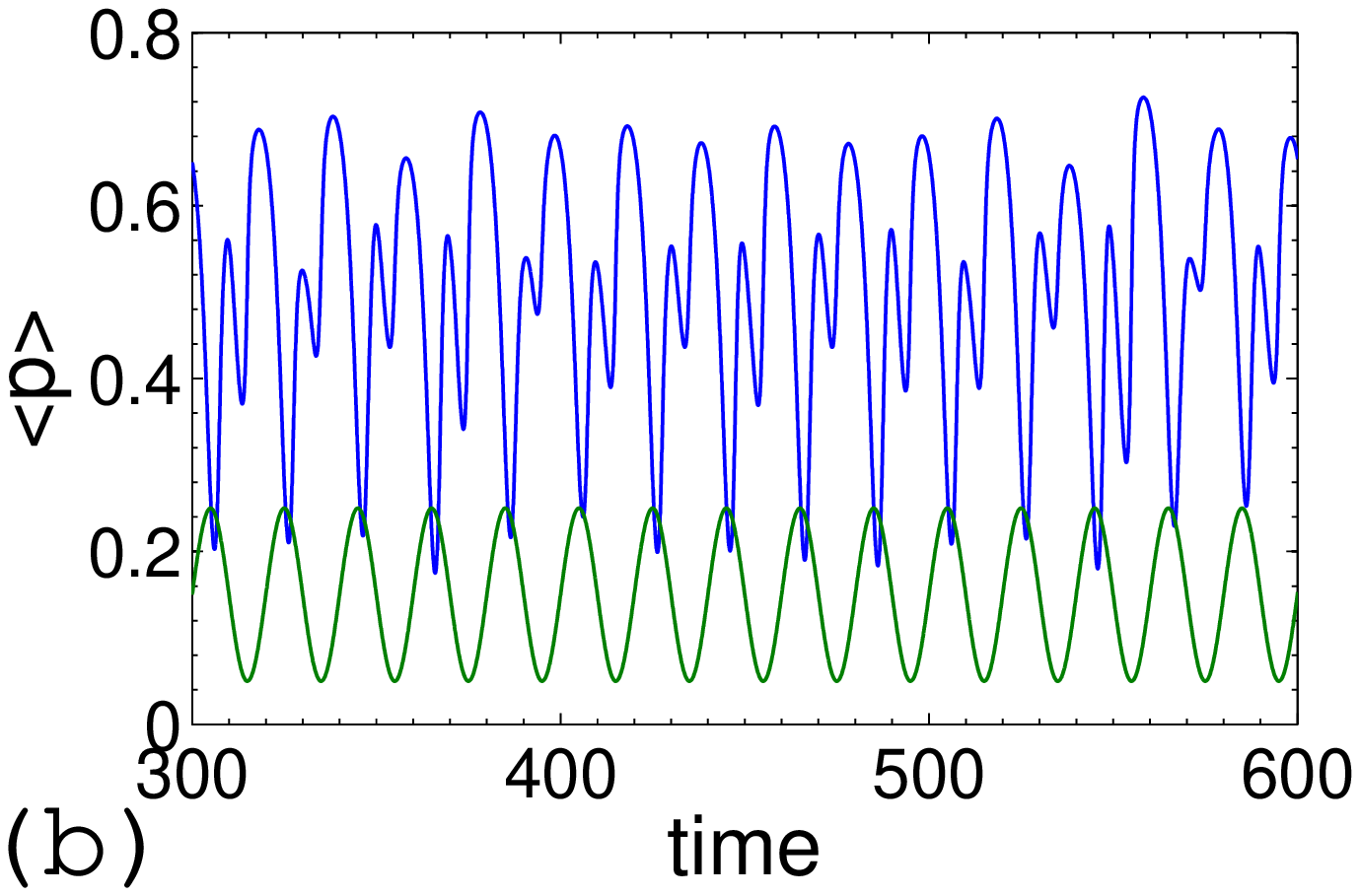}\\
%(c)\includegraphics[angle=0,scale=0.10]{picture/Fig980freenoise.eps}
%(d)\includegraphics[angle=0,scale=0.10]{picture/Fig1000freenoise.eps}
\includegraphics[angle=-0,scale=0.30]{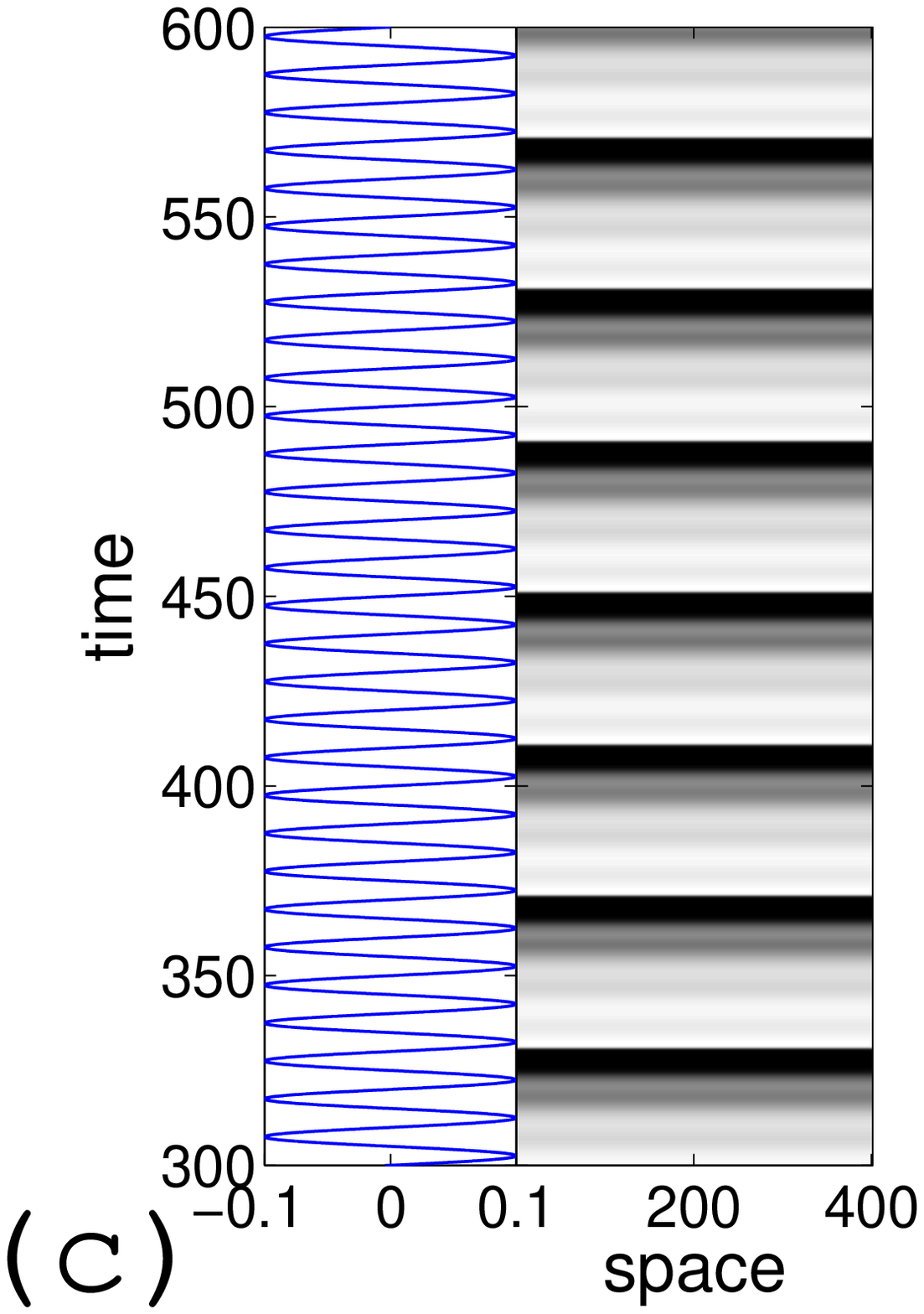}
\caption{(color) Sequences of the mean concentration $\langle
p\rangle$, the noise free. (a) The $4:1$ frequency locking
oscillation with the values of the parameters are $A=0.1$ and
$\omega=2\pi\times 10^{-1}$. (b) The $1:1$ frequency locking
oscillation with the values of the parameters are $A=0.1$, but
$\omega=\pi\times 10^{-1}$. (c) Space-time diagram (right panel)
displays the time evolution of the patterns in the one-dimensional
space. For comparison, the curve at the left panel shows the
periodic external force. In panel (c), the left and right window
share the same ordinate which is time. Note that the green lines only
denote the exact temporal period of the external forces rather than its
values to comparing.} \label{locking}
\end{figure}

Two types of frequency locking phenomena are observed in the
spatially extended model~\eqref{eq:II2}. They are $4:1$ and $1:1$
frequency locking. In Fig.~\ref{locking}(a) and (b), we have plotted
the temporal evolution of the variable $p$ when the amplitude $A$ is
above the threshold, where the $4:1$ and $1:1$ frequency lockings
take place, respectively. In these cases, the spatially homogeneous
oscillation patterns depend on the initial conditions, and two-phase
patterns with a phase shift and separated by stationary Ising front
[In particular, for the initial conditions coming from the
Ref.~\cite{medvinsky:311} that is $p(x,y,0)=p^{*}$, and
$h(x,y,0)=h^{*}+\epsilon x +\delta$, here $\epsilon$ and $\delta$
are parameters.], or alternative homogeneous oscillations come out
[see Fig.~\ref{locking}(c)]. From the random initial conditions
prepared by randomly perturbing around the homogeneous steady state
$(p^{*},h^{*})$, we obtain $4:1$ resonant homogeneous oscillations,
as an example shown in Fig.~\ref{locking}(c), which plots the
space-time figure for the homogeneous patterns to compare with the
external force, and illustrates that the patterns are well $4:1$
frequency locking. Moreover, the $4:1$ frequency locking was
demosntrated with a power spectrum as shown in
Fig.~\ref{figfrequency}, i.e.,
$T_{out}:T_{in}=\frac{1}{\omega'}:\frac{2\pi}{\omega}=4:1$. Note
that the green lines in Fig.~\ref{locking} only denote the exact
temporal period of the external forces rather than its values to
comparing.

\subsection{Dynamics under both noise and external forces}

\begin{figure}[h]
(a)\includegraphics[angle=0,width=7cm]{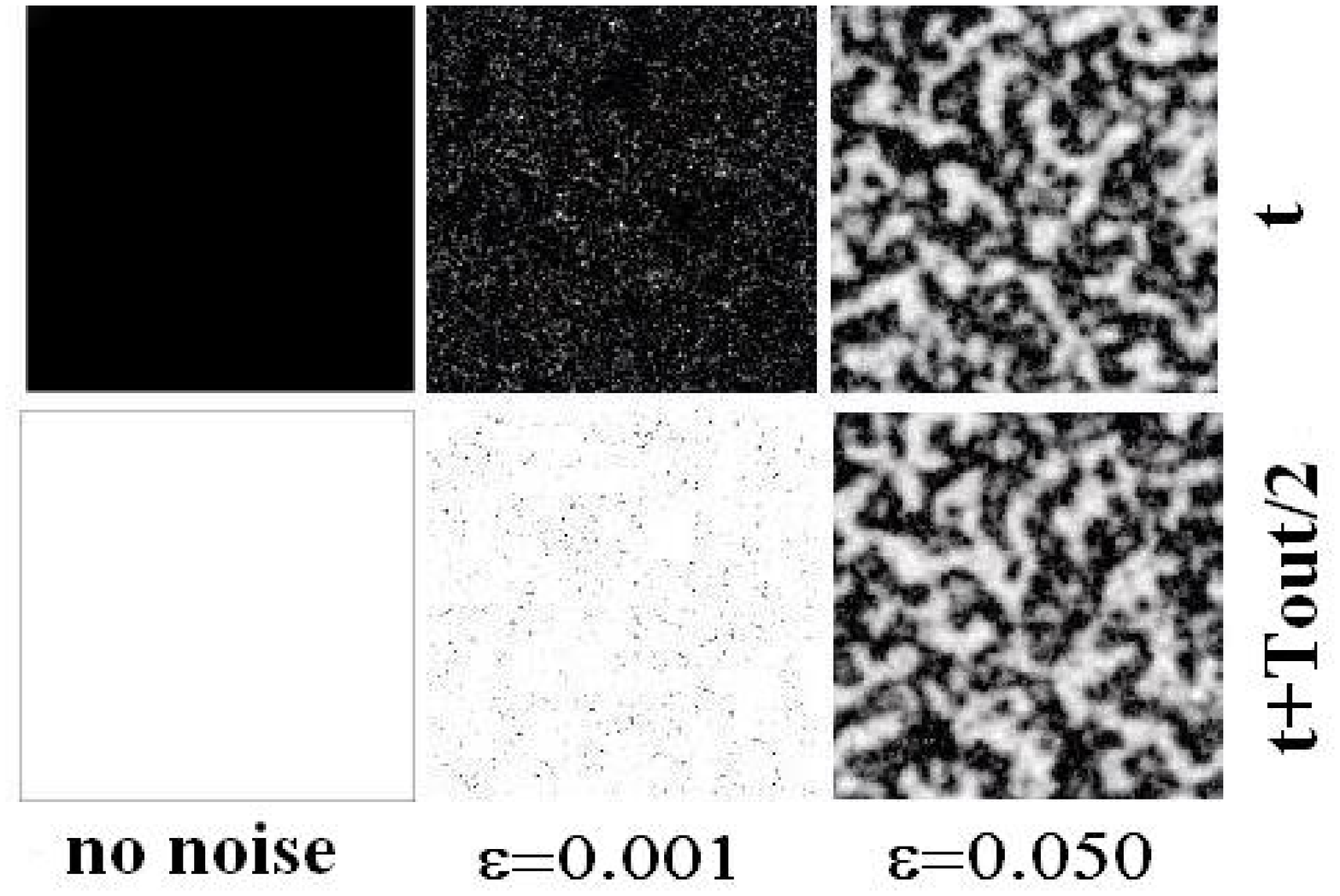}
(b)\includegraphics[angle=0,width=6cm]{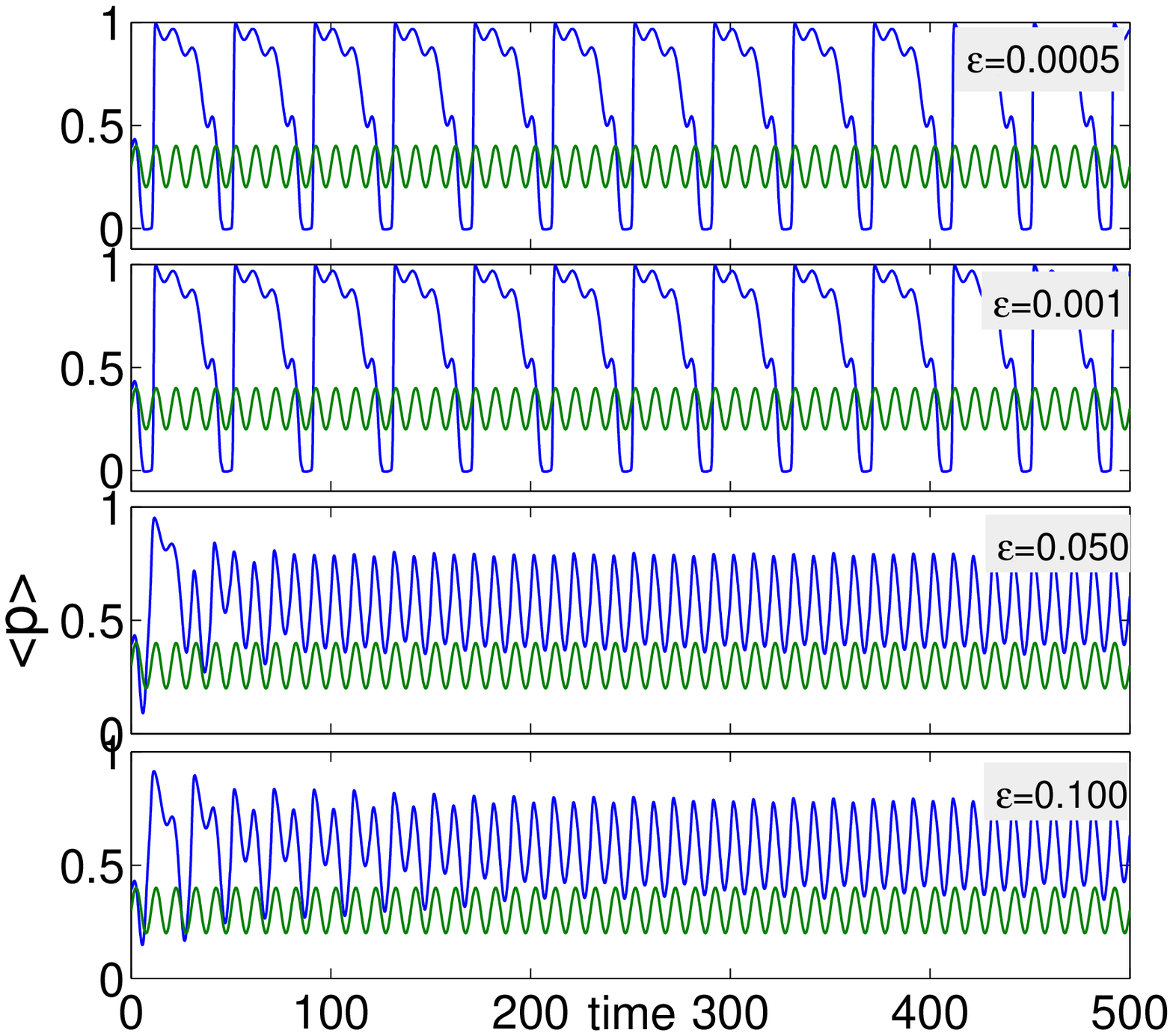}
\caption{Spatial frequency locking patterns and the time series of
the mean value $\langle p\rangle$ of the concentration $p(x,y)$ with
respect to different noise intensities $\varepsilon$. (a)
Grey-scaled snapshots of spatial frequency locking patterns with
different noise intensities $\varepsilon$, where the noise intensity
is 0.001 and 0.05 for middle column and right-hand column. (b) The
change of the frequency locking with respect to the different noise
intensities. Same situation as the Fig.~\ref{locking}(a) but
$\varepsilon$. [see the \emph{Movie-4}, \emph{Movie-5} for
Fig.~\ref{noiseandforce1}(a), and additional movies format available
from Ref.~\cite{movies}]} \label{noiseandforce1}
\end{figure}

Now, we turn on the additive noise and periodic forces in the
system~\eqref{eq:II2}. To check their effects on the
system~\eqref{eq:II2} within the $4:1$ frequency locking regimes, we
adjust the noise strength and correlation time. At first series of
simulations, we adopt $\tau=1.0$, and adjust the noise intensity,
$\varepsilon$. Considering the influences of noise and spatial
distribution on the system, we performed simulation starting at
$\varepsilon=0.001$, and then $\varepsilon$ is increased in small
steps $\Delta \varepsilon=0.001$ until the noise intensity is enough
large values. The noise drastically changes the previous scenarios
in the system including the spatial patterns and frequency locking
when the noise intensity is strong enough [see
Fig.~\ref{noiseandforce1}]. First, the spatial patterns with
homogeneous oscillation are replaced by the spatial heterogeneous
oscillations. This means that the distribution of the species may
appear spatial patterns and its bloom is periodic in the space when
some fluctuations work on the system [see
Fig.~\ref{noiseandforce1}(a)]. Second, the frequency locking ($4:1$)
shifts into the other type frequency locking ($1:1$) [see
Fig.~\ref{noiseandforce1}(b)].

From our numerical results [cf. Fig.~\ref{noiseandforce1}], one
could conclude that homogeneous oscillations depends on the noise
intensity. The oscillations are only slightly perturbed, or not
affected when the noise intensity $\varepsilon$ is small. As
$\varepsilon$ is increased, the homogeneous oscillating patters lose
its stability and associate with the frequency locking changed.
Figure~\ref{noiseandforce1}(a) (middle column) shows that
oscillating two-phase pattern with local spotted patterns will
appear at $\varepsilon=0.001$. When the noise intensity further is
increased, the spotted patterns show aggregation behaviour, and
become a large clusters in the space. Figure~\ref{noiseandforce1}(a)
(right-hand column) shows a clear spatial oscillation patterns with
a large clusters will appear at $\varepsilon=0.050$. In
Fig.~\ref{noiseandforce1}(b) we have plotted the temporal evolution
of the variable $\langle p\rangle$ when the noise intensity is taken
different values. It is interesting to realize that the resonant
patterns and the frequency locking do not alter by increasing the
noise intensity within the large interval [e.g. 0.0005 to 0.001;
0.05 to 0.10]. The variable $\langle h\rangle$ also exhibits the
similar characteristics in model.

\begin{figure}[h]
(a)\includegraphics[angle=0,width=7cm]{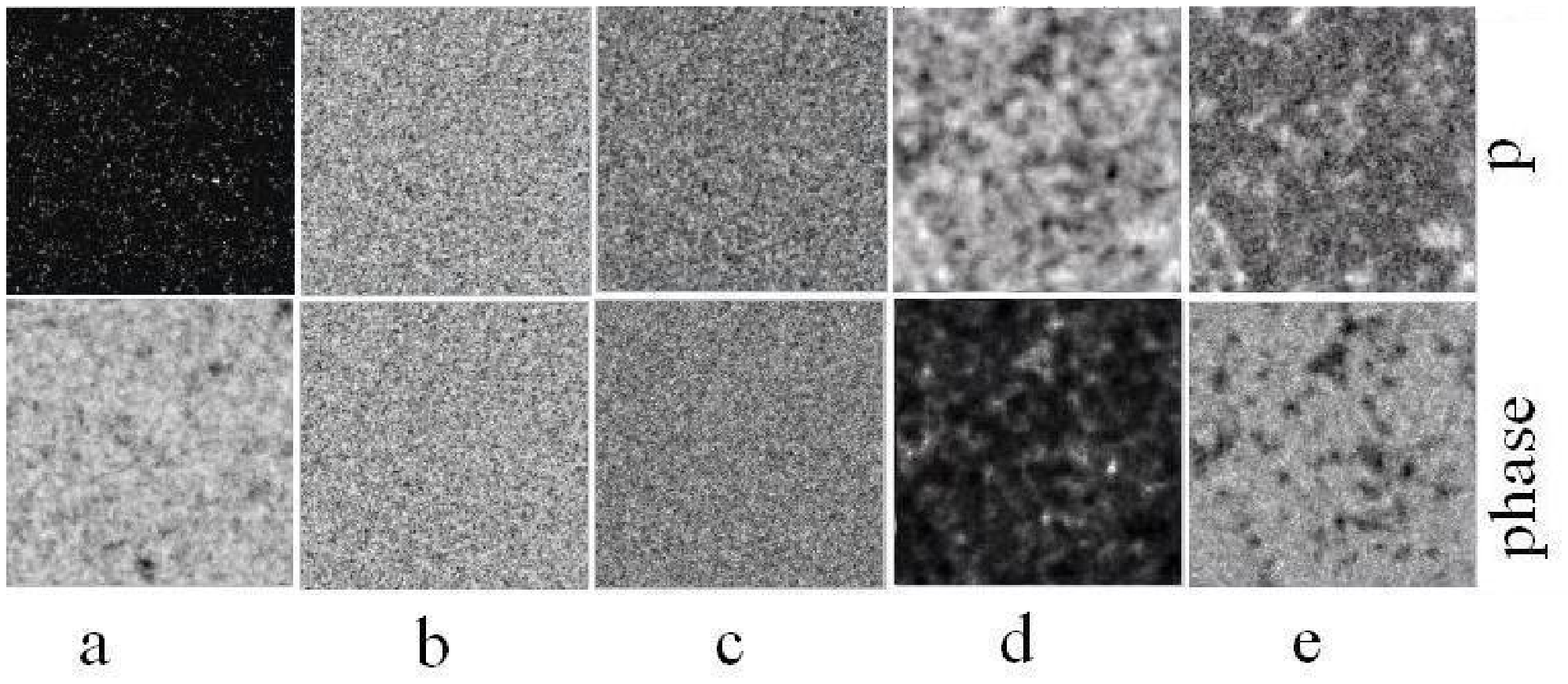}
(b)\includegraphics[angle=0,width=6cm]{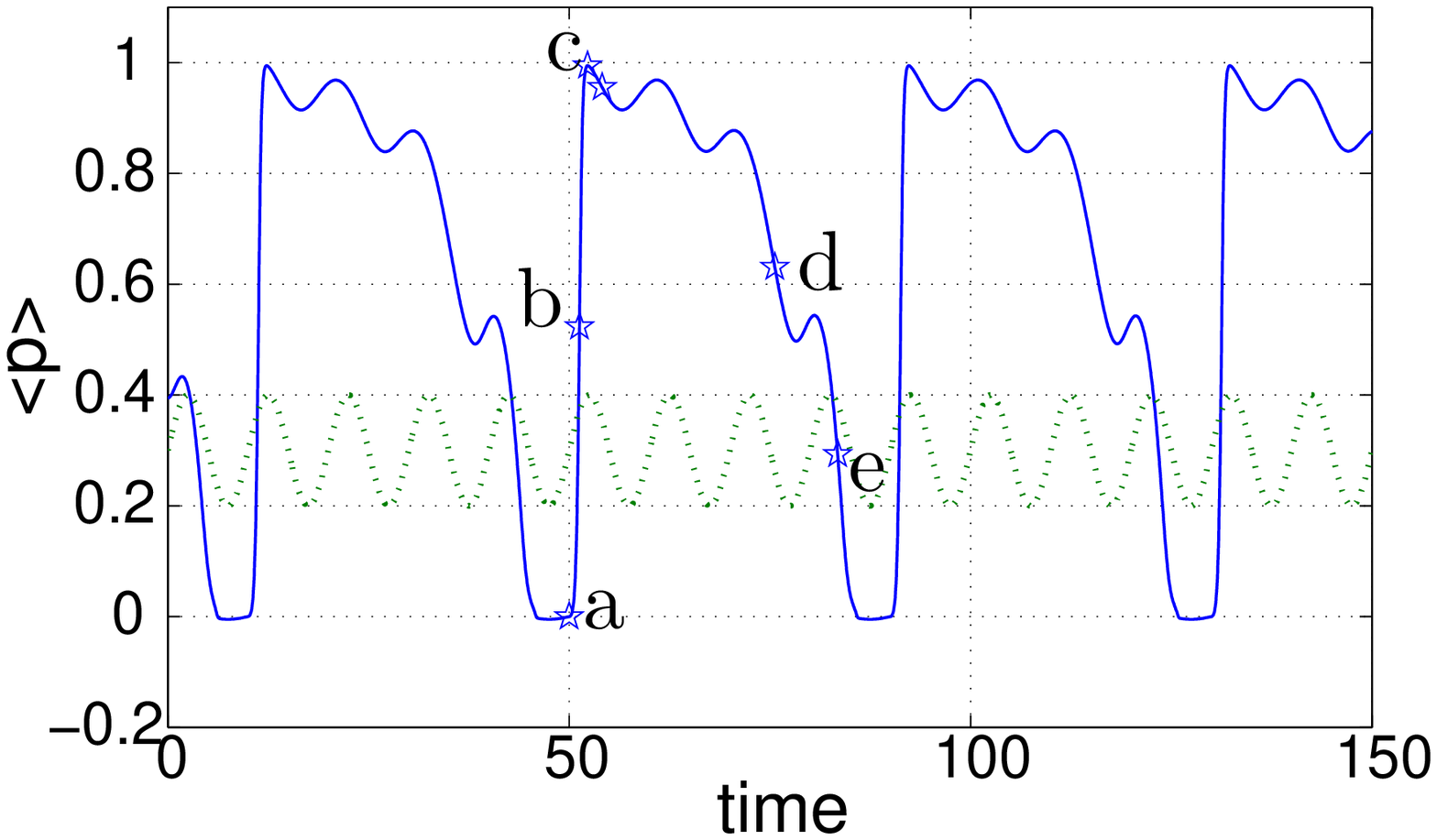}
\caption{Typical spatial pattern formation to the forces-noise
system~\eqref{eq:II2} within the 4:1 frequency locking region as in
case Fig.~\ref{noiseandforce1}(b) [$\varepsilon=0.005$]. (a)
Grey-scaled snapshots of spatial patterns of the variable $p$ and
the phase angle, $\theta$. [the time from the left to right]. (b)
shows the time series of the mean concentration $p$ corresponding to
the snapshots in (a). [see the \emph{Movie-6} for
Fig.~\ref{noiseandforce2}(a), and additional movies format available
from Ref.~\cite{movies}. Note that the movie is 25 frames per second]. } \label{noiseandforce2}
\end{figure}

In order to elucidate the evolutional processes of the spatial
structure, we have depicted typical spatiotemporal patterns of the
system~\eqref{eq:II2} within one period $T_{out}$, as shown in
Fig.~\ref{noiseandforce2} and the \emph{movie-6}. The other periods
also exhibit the similar characteristics.

 The previous Figs.~\ref{locking} and \ref{noiseandforce1} exhibit
 an aspect about the frequency locking and resonant pattern which
 have received considerable attention in the recent years, namely, the
response of the system to a periodic forces may be enhanced by the
presence of
noise~\cite{ChangsongZhou,Petrov1997,lin-2004-resonance,PhysRevLett.76.2609,sagues:829,PhysRevLett.74.2130}.

\begin{figure}[h]
\includegraphics[angle=0,width=7cm]{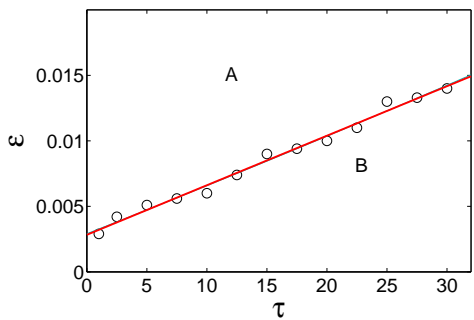}
\caption{Phase diagram in $\varepsilon-\tau$ parameter space, with
$A=0.1$ and $\omega=2\pi\times 10^{-1}$. There are two states: 4:1
region (B) and 1:1 region (A). The solid lines is a least-square fit
of circles data.} \label{phase}
\end{figure}

The role of temporal correlation $\tau$ of the colored noise is the
other significant except it intensity in inducing and controlling
the spatial patterns formation and transition of the resonant
patterns. Now, it is natural to ask what is the effect consequence
of the temporal correlation of the colored noise. Especially, the
phase diagram of the $\varepsilon-\tau$ parameter space. In order to
well understand the phase transition by the influence of temporal
correlation $\tau$, we performed a series of simulations, fixing the
$\tau$ and scanning the noise intensity, $\varepsilon$, when the
frequency locking evidently changes in the long term [here we run
the time up to 5000] and recorded the data. Figure~\ref{phase}
summarizes the results from the numerical simulations, in which the
region $A$ and $B$ are corresponding to the 1:1 frequency locking
and 4:1 frequency locking respectively. Figure~\ref{phase} depicts
the transition point of frequency locking shifts toward higher
values of the noise intensity as the correlation time is increased,
that is, $\tau$ softens the effect of the noise.

\section{Conclusion and Discussion}

It is known that, an external periodic forces applied to a nonlinear
pendulum can cause the pendulum to become entrained at a frequency
which is rationally related to the applied frequency, a phenomenon
known as the frequency-locking~\cite{Thompson}. A recent theoretical
analysis shows that an array of coupled nonlinear oscillators can
exhibit spatial reorganization when they subjected to external
periodic forces~\cite{Pierre154664}. In this paper, we present a
spatial version of phytoplankton-zooplankton model that includes
some important factors such as external periodic forces, random
fluctuations, and diffusion processes. The modified system based on
the original model by
Scheffer~\cite{Scheffer1991a,PhysRevE.64.021915}, but here it can
exhibit frequency-locking phenomena and resonant patterns. Our
results show that the noise and the external periodic forces play a
constructive role in the Scheffer's model: first, the noise can
enhance the oscillation of the species' density and format a large
clusters in the space. Second, the external periodic forces can
induce 4:1 and 1:1 frequency-locking and spatially homogeneous
oscillation to appear. Third, the resonant patterns are observed in
the system when the spatial noise and external periodic forces are
turned on, moreover the frequency-lockings transit when the noise
intensity increased. These result from outside the domain of Turing
instability. In the Appendix~\ref{app:sec}, we further provide a
simple Turing linear stability analysis with the help of the
numerical calculation by using the Maple software.

\begin{figure}[h]
\includegraphics[angle=0,width=6cm]{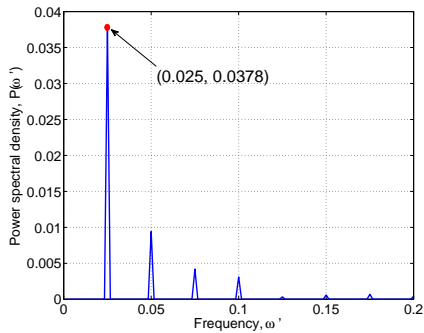}
\caption{Corresponding the power spectrum for the density of
variable $p$ as a function of its frequency with the parameters
values $\omega=2\pi\times 10^{-1}$ and $\varepsilon=0.001$ [the same
as Fig.~\ref{locking}(a)]. Note that the period was normalized to units so that $T_{out}=1/\omega'$.} \label{figfrequency}
\end{figure}

It is worth emphasizing that the frequency lockings and resonant
patterns only appear when noise and the external forces present in
the model~\eqref{eq:II2}. We here consider the cases when the
unforced system lies outside the Turing instability [see
Appendix~\ref{app:sec} for details]. The typical power spectra
$P(\omega')$ related to the density of the phytoplankton for the
oscillations within 4:1 frequency locking is analysis, as shown in
Fig.~\ref{figfrequency}. The power spectra related to the density of
zooplankton [not shown here] is similar. The presence of a prominent
and well defined peak in the power spectrum of
Fig.~\ref{figfrequency} at a nonzero frequency characterize
oscillating behavior. From Fig.~\ref{figfrequency}, one could see
that the oscillation behavior relates to the external periodic
forces and the natural frequency, $\omega_{H}$ [see the
Appendix~\ref{app:sec}] of the system. Significantly, oscillations
are enhanced in the system when the single noise term presents.
These results indicate that the oceanic plankton bloom may partly
due to the interplay between external forces and stochastic factors
instead of deterministic factors. Our results also may help us to
understand the effects arising from undeniable subject to random
fluctuations in oceanic plankton bloom.

In this paper, we study the effects of noise and external forces on
spatially extended  phytoplankton-zooplankton system in static media
(no prominent the advection term). However, in the oceanic
ecological systems the biological processes among the species are in
a fluid environment, such as turbulent flows or chaotic advection.
Recently, a few authors consider the mixing of the flow and
diffusion processes using a well-known standard model of chaotic
advection~\cite{ChangsongZhou,PhysRevE.66.066208,PhysRevLett.91.150601,
Hernandez-Garcia,Neufeld,abraham:373,hernandezgarcia-2002-12,Tzella}
in the excitable media, especially, Reigada et al~\cite{Reigada}
found that plankton blooms also can be induced by turbulent flows.
So, a further study in our work will analyze the effects arising
from the mixing of the advection, diffusion, the external forces and
noise. It is also interesting to apply the problem of oceanic
plankton bloom and the spatial structure observed by the field as a
introduction in Section I.

\begin{acknowledgments}

We thank Professor Janet W. Campbell, the Center of Excellence for
Coastal Ocean Observation and Analysis (COOA), for providing and
permitting us using the satellite images, Fen-Ni Si, China Academy
of Engineering Physics, for enlightening discussion about the
colored noise, and Professor Weiming Wang, School of Mathematics and
Information Science, Wenzhou University, for reading the original
manuscript. %Especially, Professor Bai-Lian Li encourage the
%Quan-Xing Liu work on this problem.
The authors thank Professor R.
Reigada from the University of Barcelona for pointing out
Ref.~\cite{Reigada}. This work was supported by the National Natural
Science Foundation of China under Grant No. 10471040, the Natural
Science Foundation of Shan'xi Province Grant No. 2006011009, and the
Youth Science Foundation of Shan'xi Provence 2007021006.

%The first author acknowledge helpful discussions with Prof. Bai-Lian
%Li from University of California.
\end{acknowledgments}

\appendix
\section{Stability analysis with
the help of Maple\label{app:sec}}

This appendix is devoted to the numerical analysis of unforced
system~\eqref{app:eq:II1}.
\begin{subequations}
\label{app:eq:II1}
\begin{equation}
\frac{d p}{d t}=r
   p(1-p)-\frac{ap}{1+bp}h
 ,\label{app:II1a}
\end{equation}
\begin{eqnarray}
\frac{d h}{d t}=\frac{ap}{1+bp}h-m
h-f\frac{nh^2}{n^2+h^2},\label{app:II1b}
\end{eqnarray}
\end{subequations}
Setting the left-hand sides of system~\eqref{app:eq:II1} to zero,
we obtain
\begin{eqnarray}
g_{1}(p,h)=r
   p(1-p)-\frac{ap}{1+bp}h
 ,\label{app:II2a}\\
g_{2}(p,h)=\frac{ap}{1+bp}h-m
h-f\frac{nh^2}{n^2+h^2}.\label{app:II2b}
\end{eqnarray}
The Eqs.~\eqref{app:II2a} and~\eqref{app:II2b} exist a unique
interior equilibrium $(p^{*}, h^{*})$, when parameter $f$ less than
0.445.

%With the parameters we take, the positive equilibrium as follows:
%\begin{equation}\label{app:eq3}
% p^{*}=0.3944144061, h^{*}=1.799844006.
%\end{equation}
Now we consider the stability of this positive equilibrium. The
Jacobian matrix at the positive equilibrium $(p^{*}, h^{*})$ is

\begin{equation}\label{app:eq4}
 J=\left(
\begin{array}{cc}
\Big[\frac{\partial g_{1}}{\partial p}\Big]_{(p^{*}, h^{*})} & \Big[\frac{\partial g_{1}}{\partial h}\Big]_{(p^{*}, h^{*})} \\
\Big[\frac{\partial g_{2}}{\partial p}\Big]_{(p^{*}, h^{*})} & \Big[\frac{\partial g_{2}}{\partial h}\Big]_{(p^{*}, h^{*})} \\
\end{array}
\right).
\end{equation}
Its eigenvalues, $\lambda_{1,2}$,  were listed in the
Table~\ref{tab:table1} with the different parameter values $f$. The
first type of instability is associated with a Hopf bifurcation in
the spatially uniform system. This instability will appear at the
threshold $f_{H}$ and oscillate homogeneously with the natural
frequency $\omega_{H}=\textrm{Im}(\lambda_{1,2})$. From
Table~\ref{tab:table1}, one could see that the natural frequency
equals to 0.8221 if $f=0.30$.
%\begin{equation}\label{app:eq5}
%\lambda^2-0.0946160410\lambda+0.6781376762=0.
%\end{equation}
%\begin{widetext}
\begin{table}
\caption{\label{tab:table1} The eigenvalues of the Jacobian
matrix~\eqref{app:eq4}}
\begin{ruledtabular}
\begin{tabular}{ccccc}
  $f$       & 0.30 & 0.3397 & 0.3398 & 0.3399 \\ \hline
  $\lambda_{1,2}$ & .0473$\pm$.8221\emph{i} & .0001$\pm$.8040\emph{i} & .0000$\pm$.8039\emph{i}& -.0001$\pm$.8039\emph{i} \\
\end{tabular}
\end{ruledtabular}
\end{table}

Now we consider the spatial inhomogeneous perturbation by using the
techniques of Koch and Meinhardt~\cite{RevModPhys.66.1481}.
\begin{equation}\label{app:eq5}
 \delta p=\delta p_{0}\exp(\lambda't+i\bm k \bm r), \delta h=\delta h_{0}\exp(\lambda^{'}t+i\bm k\bm
 r),
\end{equation}
with wave number vector $\bm k=(k_{x},k_{y})$, $|\delta p_{0}|$,
$|\delta h_{0}|\ll 1$ and imaginary unit $i^2=-1$. Due to the
zero-flux boundary conditions, $k_{x}$ and $k_{y}$ take only
discrete values
\begin{equation}\label{app:eq6}
 k_{x}^{n}=k_{y}^{n}=n\pi/L, n=0, 1, 2 \cdots,
\end{equation}
where $L$ is the size of the space. Each $k_{x}^{n}$ is associated with a ``frequency'' $\omega_{n}$,
which can be a complex number. The functions $\omega_{n}(\bm k^{n})$
are found by substituting expression~\eqref{app:eq5} into following
equation
\begin{subequations}
\label{app:eq7}
\begin{equation}
\frac{\partial p}{\partial t}=r
   p(1-p)-\frac{ap}{1+bp}h+d_{p}\nabla^2 p
 ,\label{app:eq7a}
\end{equation}
\begin{eqnarray}
\frac{\partial h}{\partial
t}=\frac{ap}{1+bp}h-mh-f\frac{nh^2}{n^2+h^2}+d_{h}\nabla^2
   h.\label{app:eq7b}
\end{eqnarray}
\end{subequations}
Retaining terms up to first order in $\delta p$ and $\delta h$, we
get linearized equation:
\begin{equation}\label{app:eq8}
    J'\left(
\begin{array}{c}
\delta p  \\
\delta h  \\
\end{array}
\right)=0,
\end{equation}
 with
\begin{widetext}
\begin{equation}\label{app:eq10}
  J'=\left(
\begin{array}{cc}
\Big[\frac{\partial g_{1}}{\partial p}\Big]_{(p^{*}, h^{*})}-d_{p}(\bm k^{n})^2-\lambda' & \Big[\frac{\partial g_{1}}{\partial h}\Big]_{(p^{*}, h^{*})} \\
\Big[\frac{\partial g_{2}}{\partial p}\Big]_{(p^{*}, h^{*})} & \Big[\frac{\partial g_{2}}{\partial h}\Big]_{(p^{*}, h^{*})}-d_{h}(\bm k^{n})^2-\lambda' \\
\end{array}
\right).
\end{equation}
\end{widetext}
The perturbation amplitudes $\delta p_{0}$ and $\delta h_{0}$ can be
different from zero if and only if the $\det J'=0$.

Turing instability is expected to occur for finite $\bm k^{n}>0$,
and at least one of the real parts becoming greater than zero. If
$f$ is taken as a control parameter, the critical point is reached
if the determinant $J'$ has a pure imaginary eigenvalue, i.e.,
$f=f_{T}$, $\textrm{Re}(\lambda'_{1,2})=0$. And for the critical
wave number can be written as
\begin{equation}\label{app:eq9}
   (\bm k_{c}^{n})^2=\frac{d_{h}J_{11}+d_{p}J_{22}}{2d_{p}d_{h}},
\end{equation}
$J_{11}$ and $J_{22}$ are the elements of the Matrix $J$.

\begin{table}
\caption{\label{tab:table2} The eigenvalues of the Jacobian
matrix~\eqref{app:eq10}}
\begin{ruledtabular}
\begin{tabular}{cccc}
  $f$              & 0.30                                         & 0.3124              & 0.3125 \\ \hline
  $\lambda'_{1,2}$ & -0.0704$\pm$0.8177\emph{i} & -0.0003$\pm$0.8174\emph{i}& 0.0003$\pm$0.8174\emph{i} \\
  $\bm k_{c}^{n}$      & 0.6539                     &   0.3480  &  0.3444
\end{tabular}
\end{ruledtabular}
\end{table}
Table~\ref{tab:table2} lists the eigenvalues of the Jacobian
matrix~\eqref{app:eq10} and the  critical wave number $\bm
k_{c}^{n}$, from which one can see that our simulation outside the
Turing instability regime when the $f=0.30$.

The Maple program available on request.

%\bibliographystyle{APS} %amsplain
%\newpage %Just because of unusual number of tables stacked at end
\bibliographystyle{unsrt} % halpha, habbrv, amsplain, hapalike, hacm, plainyr,
%siam, apj, unsrt, elsart-harv
%\bibliography{apssamp1}% Produces the bibliography via BibTeX.

\begin{thebibliography}{10}

\bibitem{report1}
Janet~W. Campbell.
\newblock The university of new hampshire center of excellence for coastal
  ocean observation and analysis (semi-annual technical report: Na16oc2740).
\newblock The Coastal Observing Center at UNH, 2005.
\newblock The two satellite images were obtained from NASA's Ocean Color
  website: http://oceancolor.gsfc.nasa.gov, under the ``image gallery" section.

\bibitem{bettencourt-2007-in-situ}
Lu\'{i}s M.~A. Bettencourt, Aric~A. Hagberg, and Levi~B. Larkey.
\newblock Separating the wheat from the chaff: Practical anomaly detection
  schemes in ecological applications of distributed sensor networks.
\newblock LA-UR-06-8235, 2007.

\bibitem{abraham:373}
Edward~R. Abraham and Melissa~M. Bowen.
\newblock Chaotic stirring by a mesoscale surface-ocean flow.
\newblock {\em Chaos}, 12:373--381, 2002.

\bibitem{Durbin}
E.~G. Durbin, R.~G. Campbell, M.~C. Casas, M.~D. Ohman, B.~Niehoff, J.~Runge,
  and M.~Wagner.
\newblock Interannual variation in phytoplankton blooms and zooplankton
  productivity and abundance in the gulf of maine during winter.
\newblock {\em Mar. Ecol.-Prog. Ser.}, 254:81--100, 2003.

\bibitem{May2003}
C.~L. May, J.~R. Koseff, L.~V. Lucas, J.~E. Cloern, and D.~H. Schoellhamer.
\newblock Effects of spatial and temporal variability of turbidity on
  phytoplankton blooms.
\newblock {\em Mar. Ecol.-Prog. Ser.}, 254:111--128, 2003.

\bibitem{Desharnais}
Gregor~F. Fussman, Stephen~P. Ellner, Nelson~G. Hairston, Laura E.~Jones Jr.,
  Kyle~W. Shertzer, and Takehito Yoshida.
\newblock {\em Population Dynamics and Laboratory Ecology}, chapter Ecological
  and Evolutionary Dynamics of Experimental Plankton Communities, page 221.
\newblock Advances in ecological research; v.37. Elsevier Academic Press,
  Amsterdam; Oxford, 2005.
\newblock Edited by R. A. Desharnais.

\bibitem{E.E.Popova10011997}
E.E. Popova, M.J.R. Fasham, A.V. Osipov, and V.A. Ryabchenko.
\newblock {Chaotic behaviour of an ocean ecosystem model under seasonal
  external forcing}.
\newblock {\em J. Plankton Res.}, 19:1495--1515, 1997.

\bibitem{EckartSteffen}
Eckart Steffen, Horst Malchow, and Alexander~B. Medvinsky.
\newblock {Effects of seasonal perturbations on a model plankton community}.
\newblock {\em Environ. Model. Assess.}, 2:43--48, 1997.

\bibitem{HorstMalchow}
Horst Malchow.
\newblock {Spatial-temporal pattern formation in nonlinear non-quilibrium
  plakton dynamics}.
\newblock {\em Proc. R. Soc. Lond. B}, 251:103--109, 1993.

\bibitem{benzi-2007}
Roberto Benzi.
\newblock Stochastic resonance: from climate to biology, 2007.
\newblock arXiv.org:nlin/0702008.

\bibitem{Malchow2004}
H.~Malchow, F.~M. Hilker, and S.~V. Petrovskii.
\newblock Noise and productivity dependence of spatiotemporal pattern formation
  in a prey-predator system.
\newblock {\em Discrete Cont. Dyn. Syst. B}, 4:705--711, 2004.

\bibitem{Vilar}
J.~M.~G. Vilar, R.~V. Sole, and J.~M. Rubi.
\newblock On the origin of plankton patchiness.
\newblock {\em Physica A}, 317:239--246, 2003.

\bibitem{KennethS.Johnson04192002}
Kenneth~S. Johnson, David~M. Karl;, S.~W. Chisholm, P.~G. Falkowski, and J.~J.
  Cullen.
\newblock {Is Ocean Fertilization Credible and Creditable?}
\newblock {\em Science}, 296:467--468, 2002.

\bibitem{sagues:829}
Francesc Sagu\'{e}s, Jos\'{e}~M. Sancho, and Jordi Garc\'{\i}a-Ojalvo.
\newblock Spatiotemporal order out of noise.
\newblock {\em Rev. Mod. Phys.}, 79:829, 2007.

\bibitem{Garca-Ojalvo}
Jordi Garca-Ojalvo and Jose~M. Sancho.
\newblock {\em Noise in spatially extended systems}.
\newblock Springer, New York, 1999.

\bibitem{Horsthemke}
W.~Horsthemke and R.~Lefever.
\newblock {\em Noise-induced transitions}.
\newblock Springer series in synergetics; v.15. Springer-Verlag, Berlin; New
  York, 1984.

\bibitem{Gammaitoni}
Luca Gammaitoni, Peter Hanggi, Peter Jung, and Fabio Marchesoni.
\newblock Stochastic resonance.
\newblock {\em Rev. Mod. Phys.}, 70:223, 1998.

\bibitem{PhysRevLett.74.2130}
Peter Jung and Gottfried Mayer-Kress.
\newblock Spatiotemporal stochastic resonance in excitable media.
\newblock {\em Phys. Rev. Lett.}, 74:2130--2133, 1995.

\bibitem{RevModPhys.70.223}
Luca Gammaitoni, Peter H\"anggi, Peter Jung, and Fabio Marchesoni.
\newblock Stochastic resonance.
\newblock {\em Rev. Mod. Phys.}, 70:223--287, 1998.

\bibitem{PhysRevLett.83.4896}
Alexander Neiman, Lutz Schimansky-Geier, Ann Cornell-Bell, and Frank Moss.
\newblock Noise-enhanced phase synchronization in excitable media.
\newblock {\em Phys. Rev. Lett.}, 83:4896--4899, 1999.

\bibitem{PhysRevE.61.R1001}
Bambi Hu and Changsong Zhou.
\newblock Phase synchronization in coupled nonidentical excitable systems and
  array-enhanced coherence resonance.
\newblock {\em Phys. Rev. E}, 61:R1001--R1004, 2000.

\bibitem{PhysRevLett.87.098101}
Changsong Zhou, J\"urgen Kurths, and Bambi Hu.
\newblock Array-enhanced coherence resonance: Nontrivial effects of
  heterogeneity and spatial independence of noise.
\newblock {\em Phys. Rev. Lett.}, 87:098101, 2001.

\bibitem{ChangsongZhou}
Changsong Zhou and J\"{u}rgen Kurths.
\newblock Noise-sustained and controlled synchronization of stirred excitable
  media by external forcing.
\newblock {\em New J. Phys.}, 7:18, 2005.

\bibitem{PhysRevE.67.021113}
J.~Buceta, M.~Iba\~nes, J.~M. Sancho, and Katja Lindenberg.
\newblock Noise-driven mechanism for pattern formation.
\newblock {\em Phys. Rev. E}, 67:021113, 2003.

\bibitem{wang:036210}
Hongli Wang, Ke~Zhang, and Qi~Ouyang.
\newblock Resonant-pattern formation induced by additive noise in periodically
  forced reaction-diffusion systems.
\newblock {\em Phys. Rev. E}, 74:036210, 2006.

\bibitem{PhysRevE.59.98}
Miguel~A. Santos and J.~M. Sancho.
\newblock Noise-induced fronts.
\newblock {\em Phys. Rev. E}, 59:98, 1999.

\bibitem{PhysRevLett.88.138301}
L.~Q. Zhou, X.~Jia, and Q.~Ouyang.
\newblock Experimental and numerical studies of noise-induced coherent patterns
  in a subexcitable system.
\newblock {\em Phys. Rev. Lett.}, 88:138301, 2002.

\bibitem{PhysRevE.58.4355}
A.~A. Zaikin and L.~Schimansky-Geier.
\newblock Spatial patterns induced by additive noise.
\newblock {\em Phys. Rev. E}, 58:4355--4360, 1998.

\bibitem{Riaz}
S.~S. Riaz, S.~Dutta, S.~Kar, and D.~S. Ray.
\newblock Pattern formation induced by additive noise: a moment-based analysis.
\newblock {\em Eur. Phys. J. B}, 47:255--263, 2005.

\bibitem{PhysRevLett.76.2609}
F.~Marchesoni, L.~Gammaitoni, and A.~R. Bulsara.
\newblock Spatiotemporal stochastic resonance in a $\varphi{}^{4}$ model of
  kink-antikink nucleation.
\newblock {\em Phys. Rev. Lett.}, 76(15):2609--2612, 1996.

\bibitem{sailer:056209}
X.~Sailer, D.~Hennig, V.~Beato, H.~Engel, and L.~Schimansky-Geier.
\newblock Regular patterns in dichotomically driven activator-inhibitor
  dynamics.
\newblock {\em Phys. Rev. E}, 73:056209, 2006.

\bibitem{Spagnolo}
B.~Spagnolo, D.~Valenti, and A.~Fiasconaro.
\newblock Noise in ecosystems: A short review.
\newblock {\em Math. Biosci. Eng.}, 1:185--211, 2004.

\bibitem{CarlZimmer04021999}
Carl Zimmer.
\newblock {Complex Systems: Life After Chaos}.
\newblock {\em Science}, 284:83--86, 1999.

\bibitem{OttarN.Bjornstad07272001}
Ottar~N. Bjornstad and Bryan~T. Grenfell.
\newblock {Noisy Clockwork: Time Series Analysis of Population Fluctuations in
  Animals}.
\newblock {\em Science}, 293:638--643, 2001.

\bibitem{PhysRevE.54.706}
S.~Ciuchi, F.~de~Pasquale, and B.~Spagnolo.
\newblock Self-regulation mechanism of an ecosystem in a non-gaussian
  fluctuation regime.
\newblock {\em Phys. Rev. E}, 54:706--716, 1996.

\bibitem{PhysRevLett.80.4099}
Jos\'e M.~G. Vilar and Ricard~V. Sol\'e.
\newblock Effects of noise in symmetric two-species competition.
\newblock {\em Phys. Rev. Lett.}, 80:4099--4102, 1998.

\bibitem{Giardina}
Irene Giardina, Jean-Philippe Bouchaud, and Marc M\'{e}zard.
\newblock Proliferation assisted transport in a random environment.
\newblock {\em J. Phys. A: Math. Gen.}, 34:L245--L252, 2001.

\bibitem{Spagnolo2002}
B.~Spagnolo and A.~La~Barbera.
\newblock Role of the noise on the transient dynamics of an ecosystem of
  interacting species.
\newblock {\em Physica A}, 315:114--124, 2002.

\bibitem{Valenti2004}
D.~Valenti, A.~Fiasconaro, and B.~Spagnolo.
\newblock Stochastic resonance and noise delayed extinction in a model of two
  competing species.
\newblock {\em Physica A}, 331:477--486, 2004.

\bibitem{MercedesPascual2005}
Mercedes Pascual.
\newblock Computational ecology: From the complex to the simple and back.
\newblock {\em Plos Comput. Biol.}, 1:e18, 2005.

\bibitem{Scheffer1991a}
Marten Scheffer.
\newblock Fish and nutrients interplay determines algal biomass: a minimal
  model.
\newblock {\em Oikos}, 62:271, 1991.

\bibitem{PhysRevE.64.021915}
Alexander~B. Medvinsky, Irene~A. Tikhonova, Rubin~R. Aliev, Bai-Lian Li,
  Zhen-Shan Lin, and Horst Malchow.
\newblock Patchy environment as a factor of complex plankton dynamics.
\newblock {\em Phys. Rev. E}, 64:021915, 2001.

\bibitem{medvinsky:311}
Alexander~B. Medvinsky, Sergei~V. Petrovskii, Irene~A. Tikhonova, Horst
  Malchow, and Bai-Lian Li.
\newblock Spatiotemporal complexity of plankton and fish dynamics.
\newblock {\em SIAM Review}, 44:311--370, 2002.

\bibitem{Malchow1993}
Horst Malchow.
\newblock Spatio-temporal pattern formation in nonlinear nonequilibrium
  plankton dynamics.
\newblock {\em Proc. R. Soc. London Ser. B}, 251:103--109, 1993.

\bibitem{MercedesPascual}
Mercedes Pascual.
\newblock Diffusion-induced chaos in a spatial predator-prey system.
\newblock {\em Proc. R. Soc. Lond. B}, 251:1--7, 1993.

\bibitem{Kirk1992}
Kevin~L. Kirk and John~J. Gilbert.
\newblock Variation in herbivore response to chemical defenses: Zooplankton
  foraging on toxic cyanobacteria.
\newblock {\em Ecology}, 73:2208--2217, 1992.

\bibitem{Truscott1992}
J.~E. Truscott and J.~Brindley.
\newblock Ocean plankton populations as excitable media.
\newblock {\em Bull. Math. Biol.}, 56:981--998, 1992.

\bibitem{Buskey1993}
E~J Buskey and D~A Stockwell.
\newblock {\em Toxic Phytoplankton Blooms in the Sea}, chapter effects of a
  persistent `browntide' on zooplankton population in hte Laguna Madre of
  Southern Texas, pages 659--666.
\newblock Elsevier, Amsterdam, 1993.

\bibitem{Petrovskii2001}
Sergei~V. Petrovskii and Horst Malchow.
\newblock Wave of chaos: New mechanism of pattern formation in spatio-temporal
  population dynamics.
\newblock {\em Theor. Popul. Biol.}, 59:157--174, 2001.

\bibitem{report2}
In the physical world, the assumption that the fluctuations are fast in
  comparison with the relevant systems time scales may not always be ture. To
  approximate such fluctuations by noise terms that are $\delta$-correlated
  would be inappropriate, and the reulst systems would produce inaccureate
  predictions. Here, we use the colored noise with an exponential
  time-correlated.

\bibitem{garvie:775}
Marcus~R. Garvie and Catalin Trenchea.
\newblock Optimal control of a nutrient-phytoplankton-zooplankton-fish system.
\newblock {\em SIAM J. Cont. and Opti.}, 46:775--791, 2007.

\bibitem{liuquanxing}
Quan-Xing Liu, Gui-Quan Sun, Bai-Lian Li, and Zhen Jin.
\newblock Emergence of spatiotemporal chaos driven by far-field breakup of spiral waves in the plankton ecological systems.
\newblock {\em arXiv:0704.0322}, 2007.

\bibitem{sandstede:494}
Bj\"{o}rn Sandstede and Arnd Scheel.
\newblock Period-doubling of spiral waves and defects.
\newblock {\em SIAM J. Appl. Dynam. Syst.}, 6:494--547, 2007.

\bibitem{Petrovskii2004}
Sergei Petrovskii, Bai-Lian Li, and Horst Malchow.
\newblock Transition to spatiotemporal chaos can resolve the paradox of
  enrichment.
\newblock {\em Ecological Complexity}, 1:37--47, 2004.

\bibitem{Wilson}
W.~G. Wilson.
\newblock Resolving discrepancies between deterministic population models and
  individual-based simulations.
\newblock {\em Am. Nat.}, 151:116--134, 1998.

\bibitem{misc10}
This steady state loses stability in a Hopf bifurcation to oscillations in the
  zero-dimensional mode as the parameter $f$ is decreased below a critical
  value $f_{H}=0.3397$.

\bibitem{movies}
www.iop.org.

\bibitem{sifenni}
Fen-Ni Si, Quan-Xing Liu, Jin-Zhong Zhang, and Lu-Qun Zhou.
\newblock Propagation of travelling waves in sub-excitable systems driven by
  noise and periodic forcing.
\newblock {\em Eur. Phys. J. B}, 60:Inpress, 2007.
\newblock arXiv:0706.1917.

\bibitem{Petrov1997}
Valery Petrov, Qi~Ouyang, and Harry~L. Swinney.
\newblock Resonant pattern formation in achemical system.
\newblock {\em Nature}, 388:655--657, 1997.

\bibitem{lin-2004-resonance}
Anna~L. Lin, Aric Hagberg, Ehud Meron, and Harry~L. Swinney.
\newblock Resonance tongues and patterns in periodically forced
  reaction-diffusion systems.
\newblock {\em Phys. Rev. E}, 69:066217, 2004.

\bibitem{Thompson}
J.~M.~T. Thompson and H.~B. Stewart.
\newblock {\em Nonlinear dynamics and chaos: geometrical methods for engineers
  and scientists}.
\newblock Wiley, Chichester; New York, 1986.

\bibitem{Pierre154664}
Pierre Coullet and Kjartan Emilsson.
\newblock Strong resonances of spatially distributed oscillators: a laboratory
  to study patterns and defects.
\newblock {\em Physical D}, 61:119--131, 1992.

\bibitem{PhysRevE.66.066208}
Zolt\'an Neufeld, Crist\'obal L\'opez, Emilio Hern\'andez-Garc\'ia, and Oreste
  Piro.
\newblock Excitable media in open and closed chaotic flows.
\newblock {\em Phys. Rev. E}, 66:066208, 2002.

\bibitem{PhysRevLett.91.150601}
Changsong Zhou, J\"urgen Kurths, Zolt\'an Neufeld, and Istv\'an~Z. Kiss.
\newblock Noise-sustained coherent oscillation of excitable media in a chaotic
  flow.
\newblock {\em Phys. Rev. Lett.}, 91:150601, 2003.

\bibitem{Hernandez-Garcia}
Emilio Hernandez-Garcia and Cristobal Lopez.
\newblock Sustained plankton blooms under open chaotic flows.
\newblock {\em Ecological Complexity}, 1:253--259, 2004.

\bibitem{Neufeld}
Z.~Neufeld, P.~H. Haynes, V.~Garcon, and J.~Sudre.
\newblock Ocean fertilization experiments may initiate a large scale
  phytoplankton bloom.
\newblock {\em Geop. Res. Lett}, 29:1534, 2002.

\bibitem{hernandezgarcia-2002-12}
Emilio Hernandez-Garcia, Cristobal Lopez, and Zoltan Neufeld.
\newblock Small-scale structure of nonlinearly interacting species advected by
  chaotic flows.
\newblock {\em Chaos}, 12:470, 2002.

\bibitem{Tzella}
A.~Tzella and P.~H. Haynes.
\newblock Small-scale spatial structure in plankton distributions.
\newblock {\em Biogeosciences}, 4:173--179, 2007.

\bibitem{Reigada}
R.~Reigada, R.~M. Hillary, M.~A. Bees, J.~M. Sancho, and F.~Sagues.
\newblock Plankton blooms induced by turbulent flows.
\newblock {\em Proc. R. Soc. Lond. B}, 270:875--880, 2003.

\bibitem{RevModPhys.66.1481}
A.~J. Koch and H.~Meinhardt.
\newblock Biological pattern formation: from basic mechanisms to complex
  structures.
\newblock {\em Rev. Mod. Phys.}, 66:1481--1507, 1994.

\end{thebibliography}
%\nocite{*}

\end{document}